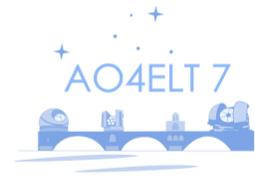

# Twenty-eight and counting

Roberto Ragazzoni[a,b,c]

[a]University of Padova, Dept. of Physics and Astronomy, vic. Osservatorio 3, 35122 Padova (Italy);
[b]INAF – Astronomical Observatory of Padova, vic. Osservatorio 5, 35122 Padova (Italy)
[c]ADONI – Laboratorio Nazionale di Ottica Adattiva

## ABSTRACT

After 28 years from the conception of the pyramid WFS several new kind of devices able to convert wavefront shape into some sort of different illumination on a detector have been conceived. While, suspending momentarily any kind of modesty, I claim credit for being among the few that contributed to show at the time that there could be much more than just a lenslet array I continued -with alternating successes- to conceive other types of such devices. I revise here only the one that I conceived on my own and, sometime stretching up and down the definition of "novel concept" I piled up 28 different WFSs, one per year from the pivotal introduction of the most successful it took to me to write down on paper.

## 1. INTRODUCTION

Dating back to 1995 several WaveFront Sensor (WFS) were available in the literature. However, for their adoption in an Adaptive Optics (AO hereafter) system, some natural selection will acts in order to make the remaining available to be countable on the finger of one hand. Actually a single finger was enough. In the years before of such a time some sort of true explorations was actually carried out. It is remarkable, for example, that shearing interferometer WFS was adopted by strategic AO for non astronomical purposes, but this was actually due to a lack of large format fast detectors, that becomes available just in the last decade of the last century. AO, in order to be of any effectiveness in the realm of the demanding, especially in the extragalactic realm, astronomical arena requires white light, broad band, detection, and artificial references was in such infancy that they were not seriously available to most if not all of the astronomical observatories. Summarizing, and with a little of irony, the only available WFS was just the Shack-Hartmann one. In this WFS the detector is located on the image plane and a lenslet array is mounted on the pupil plane. It is an ingenious development of the Hartmann test[1,2] allowing for much more light to be collected and much more further precision to be employed because of the focusing of each lens in a subaperture due to Roland Shack[3]. The Shack-Hartmann WFS was so ubiquitous that, depending upon the realm of the discussion, was used as a synonymous of WFS itself with just a few exceptions. Among these was noticeable some concepts that all have in common the use of pupil planes or leaving the pupil plane along the optical axes, moving, in the realm of concepts and ideas, in the third dimension, further to the plane that was subdivided by subapertures. The most influential author at the time was without any doubt Francois Roddier[4]. He actually developed several versions of the curvature wavefront sensor that turned out for a limited amount of time to be actually in true competition with the ubiquitous Shack-Hartmann. In fact the observations of pupil planes illumination variations was known probably from al long as the use on the sky of the telescope and I have to credit Jacques Beckers[5] for having seen, ahead of years of the time that this could become actually doable, the possibility to enlarge the field of view of an AO system. In the arena of the Authors that influenced the time I must add Brent Ellerbroek[6] and Bruce Horwitz[7]. The first for his proper treatment of several concepts in wide field, or multi conjugated, AO that become of common usage some years later, the second for two reasons: the first is that he inspired to me directly the concept of the pyramid WFS (although from a very different starting point) the second because he pointed out to me how much need for more interdisciplinarity was (and still is) needed in our world. Microscopy, telecommunication, or just shifting of a few orders of magnitude in the realm of wavelengths, led to common concepts but different terminology and different literature data bases. While this is the background where my first actual WFS has been conceived a few words are required in order to establish if a "*new*" WFS is actually *new* or it is just a slight perturbation of some existing ones. In fact counting WFSs should requires a strict definition of how much novel should be a concept to get the credit of being transformational enough. I played on this champ, I confess, in order to match the number of years from the pyramid WFS conception of this review. I played in both directions, counting variants, and assembling into a single concept what could be one or more families of different variations on the theme with different regimes in which one could play with the numbers characterizing a single concept (on the other hand, also in the case of a Shack-Hartmann, one could generalize for cases where the number of subapertures drop to one, making it a tip-tilt WFS, or goes to an equivalent subaperture much smaller than $r_0$ leaving the conventional definition just for the -restricted- realm between these two extremes) and I leave to the creativity of the reader to explore *variations on a theme* (to play a tribute to the late Roddier, with whom I had the privilege to discuss several characteristics of WFSs).

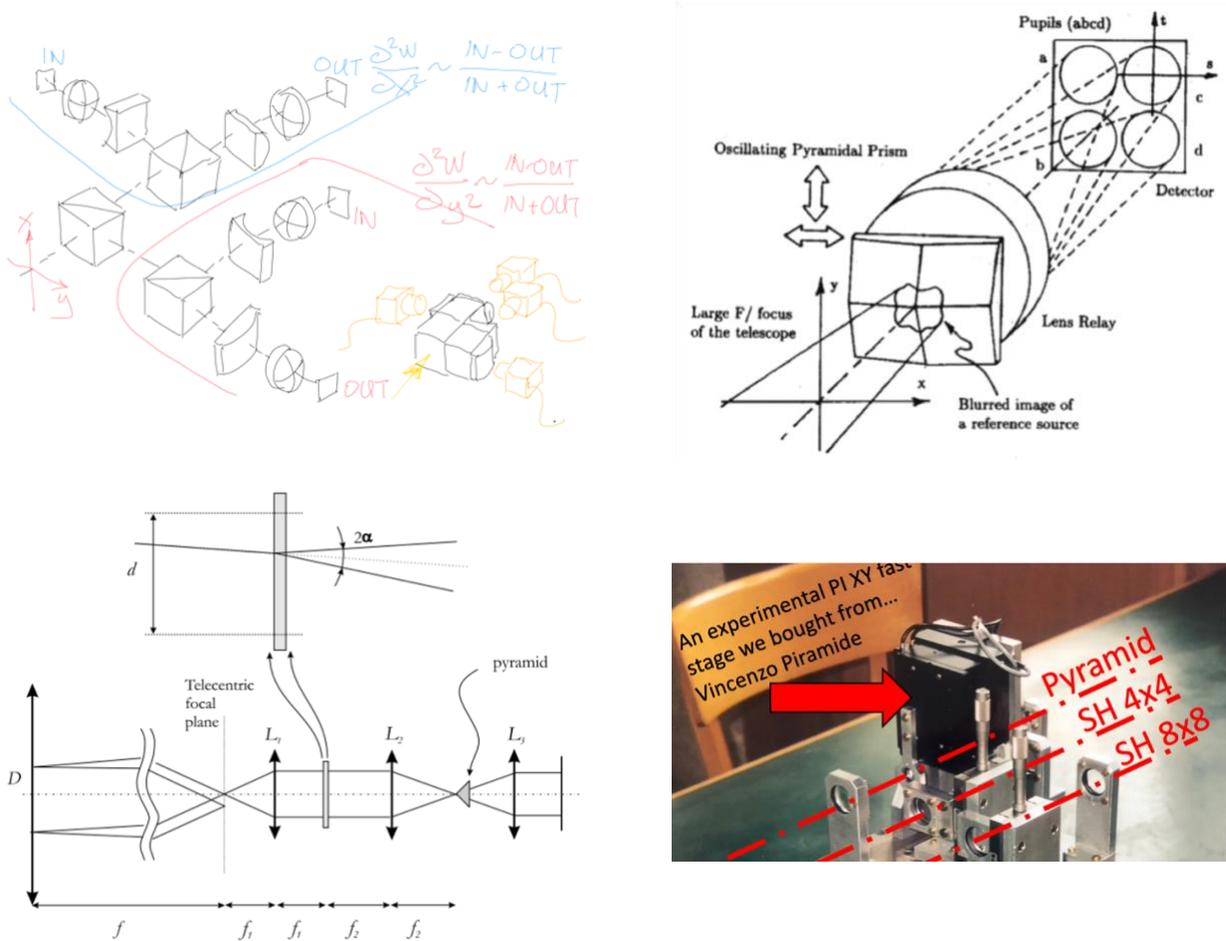

**Figure 1 - Clockwise from the upper left: A curvature WFS where the second derivative is introduced separately onto two orthogonal axes. This is WFS#0 in the text. The pyramid WFS, or WFS#1, as depicted in its seminal paper, ref.8. A way to introduce an optically equivalent modulation without any moving part, as depicted in ref.10. Finally, lower right box, the Pyramid WFS actually implemented on TNG, included two Shack-Hartmann channel that were never exploited on the sky as all resources were devoted to the pyramid one.**

## 2. STARTING FROM ZERO

Actually the curvature WFS as an alternative to the Shack-Hartmann one was such a refreshing novelty that we explored several approaches in the same framework. In fact there is one kind of WFS that was never published at the time (although later has been depicted at a conference) that offers a curvature signal for each axis of the pupil. The curvature WFS exhibits a signal proportional to the Laplacian of the wavefront. This lead to some unseen modes so that measuring the derivative of the wavefront on the edges become essentials, especially for the astronomical AO of the time where the compensation was limited and the relative importance of the very first mode played an higher role than what is today. Following the idea of placing the detectors in the exact pupil plane and changing optical power in the focal plane (the famous Roddier's modulator) only in one axis allow for this kind of WFS. We never really exploited this in detail although I can say that suffer of splitting the light into two before sensing and that suffers of the same issue of the curvature WFS, that is begin more sensitive than WFS offering the first derivative, to the noise effects. The actual implementation with cylindrical lenses also do not give the option of an automatic gain control but one could envisage a cylindrical variable optical element to deal with such a possibility. Anyway, this WFS never become reality nor it is expected to have characteristics such to be of any particular interest in the today arena (but, who knows…??) and is here numbered as "zero". So, finally we arrive to the pyramid[8] WFS. The only reason it was introduced at the time was just to change the pupil sampling with just on-chip binning. In fact we produced at the time some movies (still available on public video channels) that show the morphing of the pixels from a Shack-Hartmann to a Pyramid. This was introduced in the AdOpt@TNG, the AO module for the Telescopio Nazionale Galileo that, by a chain of events, fall under my responsibility. In this module a WFS with three options was equipped with two lenslet array to cover the 4x4 and 8x8 sampling of the pupil and a pyramid WFS (with the pyramid actual oscillating, still the most efficient in terms of photon consumption, approach). Only more than four years later, we realized the higher sensitivity of the pyramid[9] with respect ot he Shack_Hartmann one. This story would deserve a paper by its own but surely one of the contributing cause for this was just simply that everyone was a-priori skeptical on anything different than the WFS of the moment. The need for the modulation was initially seen as such a complication to avoid to use it at all but we immediately pointed out ways to introduce and eventual optical modulation[10] and, much later, we realized that such an approach could be used to engineer an equivalent modulation with an explicit chromatism[11] to augment the effectiveness of such a WFS. We actually realized in a somehow heuristic manner that no modulation at all[12] could make the WFS still a very interesting one, at the point that this option deserved the status of "different" WFS from other authoritative collegues[13]. While this made the counting still arriving to three different kind of WFSs (a pyramid with mechanical, optical, or no modulation) I decided to give the status of a fourth concept the approach of modulating during the calibration (taking advantage of the linearity) and then to close the loop with no modulation at all. This is the approach used in MAD[14,15], approved well before that a single pyramid has been tested on the sky[16] at the TNG. But MAD exploits a different approach that deserve a new chapter.

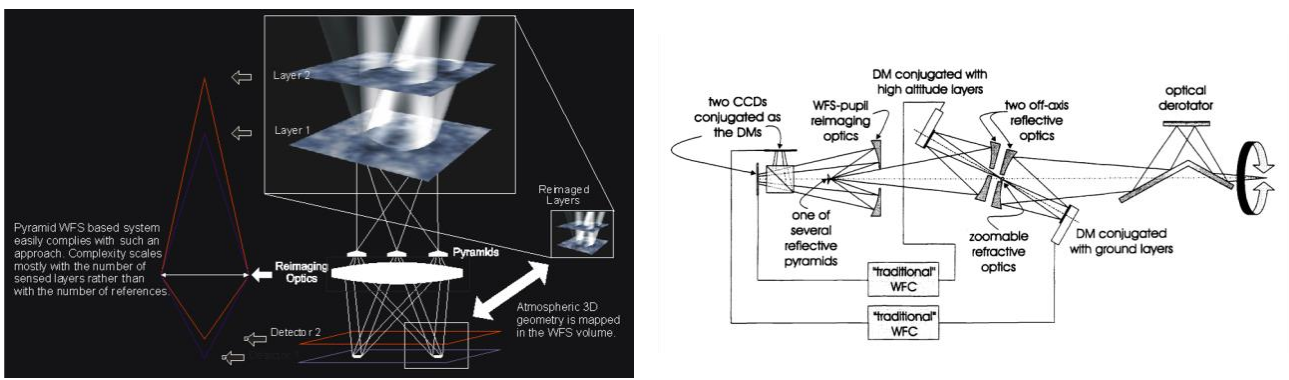

**Figure 2 - Left: the original drawing for explaining the "layer-oriented" WFS approach. One of my preferred way to see it is that in the WFS a 3D anamorphically reduced copy of the atmospheric volume to be sampled is recreated where one can properly place a detector to sample WFS as it occurs on a specific layers in the atmosphere. Right: the first proposed sketch of how to implement it in a real system.**

## 3. MULTIPLE REFERENCES, MULTIPLE PYRAMIDS

The pyramid WFS implement much of the properties that was key difference of the curvature with respect to the Shack-Hartmann, but retained the outgoing signal as the first derivative into two orthogonal axes. This made it more advantageous to implement as it could take the whole plethora of algorithm and architecture developed for the Shack-Hartmann and, moreover, exhibits a better sensitivity behavior and rejection of noise., What one would like to have more than a pyramid. Easy: two three or more pyramids. The limitation of the isoplanatic angle of AO was going to meet an end, not only because we have shown this was feasible[17] using real data collected at the telescope, but also because times become mature to implement it on a real wide field AO system. The pupil plane concept can be generalized and instead of looking just at what happen on the immediate vicinity of the entrance pupil, coupling multiple pyramids would allow for an inherently optical solver of the tomography problem. This happen at a time where WaveFront Computers was a serious issue and Read Out Noise of fast CCDs, although much better than what was available before, made a "window of opportunity" for the concept of the layer oriented approach[18]. In this approach, in fact, the four beams split by each pyramid whose pin is conjugated to the image of a reference stars, are coupled in a way that they form a three dimensional, anamorphic (the angles between the beams are augmented because of the Lagrange invariant, so that the beams looks compressed along the z-axis in a quadratic manner rather than linearly as the size of the beams) copy, where one can place detector properly conjugating to any height. In fact, whenever the reference stars would become numerous, the turbulence sensed at equivalent altitudes much different from the ones where the detector is place will appears severely smoothed out. This, not only allow for sensing of the "right" turbulence but, provided the stars form an ensemble encompassed in a certain Field of View, make an optimal filtering, allowing low order aberrations coming from other layers to be properly sensed and possibly compensated, using individual control loops. In other words the optics is an inherent part of the optical computing "diagonalizing" the interaction matrix and making this a sort of two stage real-time computer where the first part is accomplished in an optical manner. But all these details would become clearer much later, at a stage where the power of computer, following the Moore's law, made this feature less attractive, as the optical superposition of the light from several stars becomes less and less appealing with the introduction of extremely low Read Out Noise detectors (like the $L^3$CCD, for instance). As usual, the main focus at the time was how to make these real devices and the same Lagrange invariant poses apparently fundamental limits to the compression of the beams and the usability on the low size fast read out detector of the time. This, however, has been brilliantly solved through the use of individual beam compressors, as it would be the right way to name it, but for same reasons, we opted to use their symmetric behavior to "enlarge" each star individually into a commonly place focal plane, with the scale of each star being different than the scale of their relative position, as it is often in the pictorial description of the Solar System, where planets are at the right relative distance from the Sun and their size is properly scaled between each other, but with two rather different scales to avoid almost empty visualization. That has been the birth, among the others, of the so-called "star enlargers"[19] employed at the already mentioned demonstrator built for the VLT[14,15].

This unique combination of optical and numerical computing succeeded to achieve scientific results in a handful of nights where the prototype of the WFS has been successfully used onto different asterisms. It is noticeable that the interaction matrix, and hence the time required for calibration, does not depends upon the position and brightness of stars but only up to the layers adopted. In a remarkable series of nights under the wonderful Chilean nights toward the end of september 2007, unexplored portions of color magnitude diagrams of a dwarf galaxy[20] or of a globular cluster[21], and new upper limits on neutron stars characteristics[22], were identified and led to the first pyramid-based scientific publications

The filtering concept and the Fourier approach to the measurement of turbulence triggered a further elaboration on the layer-oriented approach, where specifically each layer is sensed with an optimally chosen field of view[23]. The realization that the ground layer, being the more important and also the thinnest of the ones usually encountered in the observation, could be dealt with stars in an annulus so distant from the optical axis not to be useful for higher height sensing, led to the conviction that this technique were destined to achieve good performances with a sky coverage approaching the 100% even for the regions closer to the galactic poles. The technique undergoes a number of analytical and numerical simulations[24] and the actual assessment of the performance, depending ultimately upon the desired Strehl in a certain band, moved with time as the increasing performances on bluer wavelength set a sort of moving target where the overall high efficiency of multiple pyramids over multiple Field of Views, under non exceptional conditions, become clearly not enough at a certain point.

# 4. THE FIRST LASER CONVERSION

While the new demanding performances has been the reason for the first "serious" conversion to use artificial references, one can date back to about the same date when the pyramid WFS itself has been introduced, the first recognition that using the same kind of WFS conceived to employ a natural, unresolved, reference star, is such a limiting approach that several other solutions are surely available. In a first essay at the turn of the century[25] I described a sort of generalization of the pyramid WFS that span from a crude approach in which one gives up from the beginning to measure the wavefront derivative along the elongation of an artificial reference, to the superpositions of several pyramids in order to use several portions of an artificial reference. One should recall that at the time was pragmatic to think to a WFS conceived to deal with a Rayleigh beacon rather than to a mesospheric Sodium resonant scattering one, just because the latter were hardly available at the time. But these were just sort of "*gedanken*" experiment, although I admit I was heavily influenced from the findings, circulating in the OWL environment of the time, that LGS elongation was less than a serious issue as limiting measurements just orthogonal to the elongation led to a surprisingly good enough reconstruction of the whole wavefront.

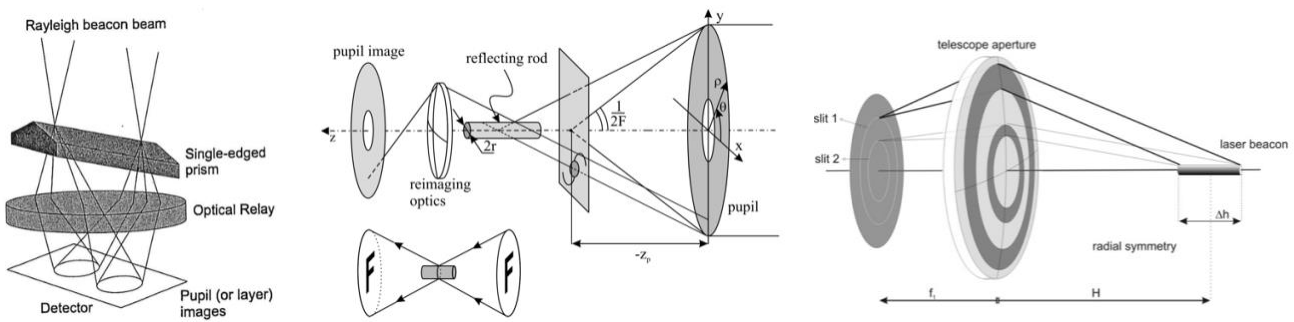

**Figure 3 - From left to right. The first description of a prismatic device to encompass the whole length of a laser reference, although limited to the sensing of the derivative along a single direction; the description of a pure z-invariant WFS encompassing a reflective rod; the insertion on the previous concepts of radial slits allowing for an effective gating leading to the measurements of the whole wavefront, although without using the whole available light.**

The first serious approach popped out with the so-called class of "z-invariant" WFS I introduced later[26]. The idea behind this approach is simple. If one find a way to sense the wavefront using optical devices that exhibits a shape that is invariant along the optical axis (the so-called z-invariance) the system would measure the wavefront using the whole column of light rather than a limited fraction of it, selected by time gating or other ways. This approach, furthermore, avoid the need for pulsed lasers that, unavoidably, takes with themselves a number of problems that at the time looks unsurmountable: each individual pulse was so powerful that all the optical trains have to dealt with, and making saturation at the Sodium layer much more severe of what was already at the time. The risk was that the dream of an artificial reference could be seriously hampered by such an inefficiency in its generation that one has to deal with a flux comparable to the one used in the early days of natural guide stars adaptive optics.

The actual adoption of a Laser Guide Star (LGS hereafter) fired from behind the cage of the secondary mirror lie at the concept of some variations on the theme of the use of a reflecting rod in the focal volume where the various height at which the LGS scatter back some light are conjugated. It is easy to form a signal proportional to a tangential derivative of the wavefront, leaving the radial derivative unknown. Use of multiple references, eventually via a sweeping mechanism, possible given the pupil plane nature of these classes of WFS, would solve the problem of global wavefront reconstruction, at least potentially. In alternative variable reflectivity along the length of the rod (perturbing the z-invariance principle, to be fair…) or by gating the light through circular slits into an intermediate plane, would gives some information useful to fully recover, at least in principle, the whole wavefront. Of course such approaches would diminish the overall efficiency, at least because not the whole light is being used. Systems that splits the light into complementary beams, although possible in principle, would requires some very complicated optomechanical systems, that are in contrast with the elegance of the z-invariant approach and have not been attempted. It is remarkable, however, that on-sky experiments of this kind of WFS (under the nickname of Pseudo Infinite Guide Star…) has been carried out atop the Canary Island using the GHRIL facility of the WHT[27].

# 5. NATURAL AGAIN

In spite, or maybe just because of, the on-sky experience with LGSs, I turned back my attention to more efficient usage of natural photons, or toward techniques to use most of the references available. The multi-pyramid systems, although elegant in principle, proven to imply a large number of mechanical axes, making some WFS the largest in the world and the ones with the highest numbers of degree of freedom, not necessarily carrying with them a proportional larger amount of effectiveness. This led to the concept of the "onduline" where a sinusoidal gratins is placed in the focal plane of a starry field and further projection of the overall field onto a pupil plane would give a signal proportional to the derivative of the wavefront along the direction orthogonal to the sinusoidal lines[28]. You would need a couple of these to achieve full wavefront reconstruction and for a complete random disposition of the stars a square root gain with the field of view would be achieved, leading to a proportionality to the field of view diameter. As one could reconstruct three dimensionally the turbulence in a layer-oriented fashion this is not so bad, but a small perturbation of the sinusoidal pattern, tailored to the starfield under observations, could remove the square root proportionality leading to a quadratic gain with the diameter of the covered Field of View. I am afraid that such an approach, although rather simply, is never been really attempted into the sky, but maybe it will be one day, eventually with some variations on the theme.

A further refinement of WFS, that could contemplate a revival to the Shack-Hartmann approach, is the one developed as hierarchical WFS[29]. In fact one of the key things that were gaining momentum in that years was the fact that the pyramid WFS takes advantage of the closed loop condition. While it is relatively easy to describe what happens when full compensation is achieved, partial compensations could lead to different degree of results[30], leaving some space even for a Shack-Hartmann WFS where the lenslet sampling is suboptimal, but taking the advantage of a relatively smaller spot because of higher order corrections carried out by a finer sampling Shack-Hartmann, or a pupil-plane (like a pyramid one) taking into account that advantage. The system could also be seen as a sort of booster mechanism in which in some parts of the closure of the loop one channel of this multiple arms WFS is more important than the others.

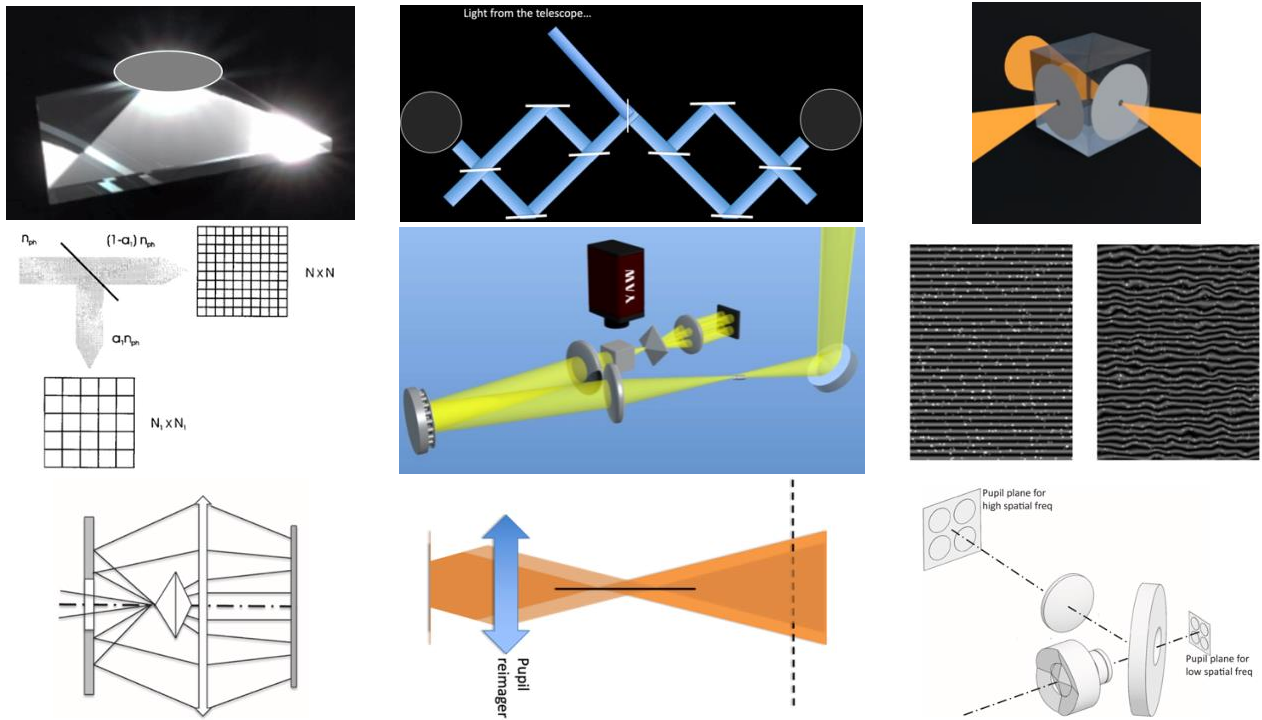

**Figure 4 – From top to bottom, each line from left to right: a pyramid with an occulting disk; a double arm interferometer as a dark WFS; a double Smartt dark WFS; principle of a hierarchical WFS; a Very Linear WFS; straight and perturbed "*onduline*" WFS; an on-axis arrangement for a multiple spatial WFS; a wire LGS-dark WFS; a double arm folded arrangement for a multiple spatial frequencies WFS.**

While this approach has not being further studied or developed in detail, some more attention has been given to what deserved the name of Global WFS[31], where simply not only the ground layer but the whole turbulence is expected to be reconstructed with a technical (where the sensing is achieved) Field of View, much larger than the scientific one. With modest apertures this would not exhibits any improvement at all, as enlarging the Field of View will makes decoupled beams simply to add unnecessary information, but with the larger apertures of the extremely large telescope in sight (from the 100m of the OWL to the 40m class of the European ELT) a significative gain can be achieved, and the extra, unnecessarily information is not needed and can be easily discarded by an extreme accurate knowledge of the lowest altitude layers. This approach would be hard to achieve with an optical layer-oriented approach, but in the meantime the read out noise of detectors lowered so much that a numerical layer-oriented would be preferable and the idea of encompassing a technical, ten arcmin field of view where plenty of relatively bright stars are available, becomes very attractive. These should be exploited with some sort of locally closed loops where the signal is retrieved from the residual uncompensated wavefront and the actual shape of the local DM, in a concept of what has been nicknamed "Very Linear WFS". As it makes sense only on very large diameter aperture this concept, although studied in relatively great detail[32,33,34,35], is apparently not being pushed forward, but I can easily predict that when larger apertures will be more numerous and more available than what is today (and in the coming decade) there could be a revamping of such an approach, unless some other technique (including an extensive use of LGSs) would dominate the arena.

Discussing with colleagues often the question is raised upon if the non modulated pyramid is the ultimate WFS (at least under some conditions that includes the possibility to close the loop in a very efficient manner) or if there are some unexplored approaches that would led to a better, or much better, capabilities, without exploring other variables (like exploiting a larger Field of View, actually taking advantages of further references, as is in the Global MCAO). Maybe during this discussions, and how to "beat" the Poissonian photons limit the idea that the perfect WFS is the one that leave no photons as outputs, popped out. This is well explored realm in other areas[36], and some ideas with a pyramid partially obscured in the central region to sense more efficiently the tip-til term, were already circulated. Specifying which is the nature of the signal such an ideal WFS would produce and then reverse engineering it has been the key thing behind a stream of ideas that get the nickname of "Dark" WFS[37] (as when the loop is actually closed, they should ideally exhibits the absence of photons to be detected) with further variations on the them by spatially filtering on the focal plane the proper signal[38] prior to injecting into pyramid-like WFSs to be used in combinations.
Further variations on this and other themes have been discussed and I like to recall a sort of "wire WFS" where the focal plane volume is interested by a physical non reflective, blocking light, "black" wire aligned with the optical image of a LGS reference[39], a sort of short circuit between various ideas, just enunciated, with little or no efforts available for their systematic description, and awaiting optimization and deep analysis.

## 6. THE SECOND (ONGOING) LASER CONVERSION

After remaining dormant for almost a couple of decades, the solution conceived for a Rayleigh LGS at the turn of the new millennia deserved further attention. The kicking factor here, however, come from a simple practical issue. In the ELT in order to deal with the elongation of LGSs at the edge of the pupil (the LGSs are fired from the side of the primary mirror, in contrast with several solution that rely on their firing from behind the cage of the secondary or, in general, from the central obstruction) a very large format detectors is required and its development could jeopardize the feasibility even of very conventional Shack-Hartmann WFS, unless a severe LGS truncation is accepted. One should note that in all these cases it would be hard to get significant information for the derivative aligned to the apparent elongation, making pupil plane WFS an election choice.

With this target in mind we conceived the so-called "INGOT" WFS, where the name just resemble (somehow vaguely) the way the key optical device is built and we do not know of any attempt to form an acronym (although it could be an interesting idea). A different way to see this WFS is to imagine to have a SH-WFS and to imagine how one would like to subdivide the LGS projected onto the focal plane to sense it efficiently in terms of usage of pixels. Any elongated beacon could be subdivided into six pixels arranged into a 2x3 grid where all the pixels are used to sense the derivative orthogonal to the elongation and the 4 pixels to the ends would gives the proper information from the two edges to collect information about the other derivative term. Of course this arrangement is variable across the pupil pane as the "central" pixels would be almost non existing for points over the pupil close to the location where the LGS is being fired, and would dominate on the opposite side. This feature is "automatically" accomplished by the ingot approach, in the same way as in the pyramid, as there is just a single cross, there is no any issue as in the SH-WFS where one has several spots and if you want to optimally use a grid of pixels you have to carefully aligned each spot with the cross among a set of four pixels per each subaperture. This consideration is behind the concept that the Ingot-WFS is the extension of the pyramid to a source that deploy in three dimensions above the observatory[40], as the LGS actually is. One is also warned that this approach gives up, by design, to use any feature due to variations of the Sodium profile along the vertical propagation to further get information on the "bad" signal. Furthermore, as the edges are, in general, a little portion of the whole LGS, and the distribution of Sodium density is non symmetric, one is tempted to further give up other sources of information, making the WFS simpler and simpler. In its simpler approach, with only three equivalent pixel, one would measure the derivative of the wavefront along the elongation just using what is usually the sharpest edge. As clouds in the sky, in fact, often the Sodium profile exhibits an abrupt change in its lowest altitude portion rather than in the highest altitude one, so giving to the overall solution a sort of best use of pixels in exchange for suboptimal performances. The apparent elongation of the LGS beacon is also continuously changing, leading to various possible options to deal with, ranging from an array of ingots to anamorphic zoom optical relay (where only one axis change in focal length) to sub-optimal solutions where the ingot is tilted to arrange for different elongation at the expense of a moderate defocus and hampering of the signal.

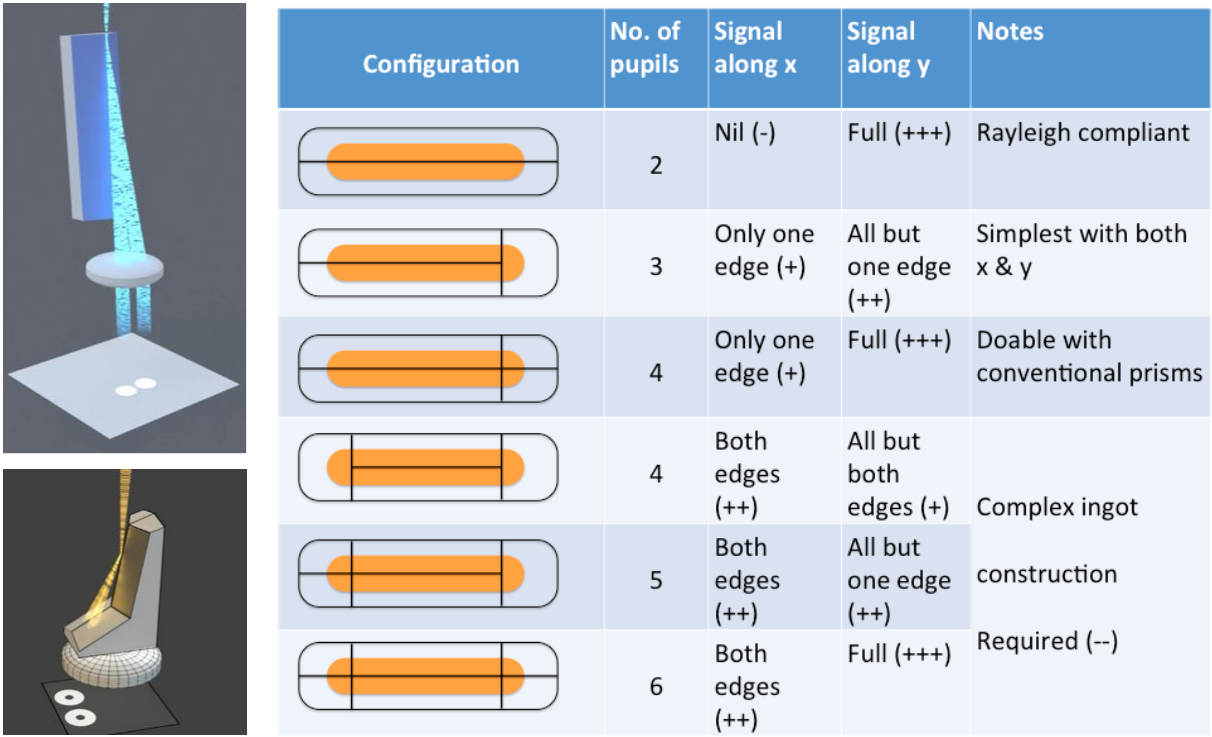

**Figure 5 – Right side: a table showing the possible arrangement of the various faces for an ingot-like WFS. Upper left: an exemplification of a 3-sided ingot-WFS. Lower left: an exemplification of a 6 faces ingot-like WFS.**

Optical solutions for ELT-class telescope on conventional format detectors with 3 and more sided ingots have been designed and analyzed analitically[41], simulated[42] and tested on optical bench of various nature[43,44], including the need to provide from the pupil plane signal to guide the ingot to place in position during an acquisition phase[45] that, assuming the use of a Sodium LGS, would change from pointing to pointing as the actual range of the beacon would depend upon the actual elevation of the telescope. While end to end simulation of some nominal design have been investigated for ELT-class telescope[46] we are still proactively awaiting the opportunity for on-sky tests.

## 7. (PROVISIONAL) CONCLUSIONS

To achieve the 28th (and -momentarily- the final one in this hypothetical -tailored- numbered list) WFS I have to add a very particular WFS conceived time ago but published only recently. It is a very custom WFS to be used just for planetary imaging and it would use the edges of a relatively bright planetary disk (like Mars or Jupiter) to collect information about both the derivatives of the wavefront using edges on the proper portions of the planetary disk. As their apparent size would change within their journey around the Sun, some variable optical device or mask should be employed. The interested reader is invited to look at the proper reference[47] but it is here worth noting that given a very specific case to deal with a very dedicated AO system, possible custom solutions could be achieved (and one could just think to Solar AO, that is by far beyond the limits of this paper).

| # | Description | Wide Field | LGS | On sky | Reference |
|---|---|---|---|---|---|
| 0 | Double curvature | | | | This paper |
| 1 | Pyramid (modulated) | | | Yes | 8 |
| 2 | Pyramid (optical modulated) | | | | 10 |
| 3 | Pyramid (non modulated) | | | Yes | 12 |
| 4 | Pyramid (modulated only to calibrate) | | | AdOpt@TNG & MAD-VLT | 15,16 |
| 5 | Layer oriented | Yes | | | 18 |
| 6 | Multiple Field of View, layer oriented | Yes | | LN-LBT | 23 |
| 7 | Double side prism | | Yes | | 25 |
| 8 | z-invariant reflecting rod | | Yes | | 26 |
| 9 | z-invariant variable reflecting rod | | Yes | | 26 |
| 10 | Radial gating z-invariant reflecting rod | | Yes | WHT | 27 |
| 11 | Multiple rods layer oriented | Yes | Yes | | 27 |
| 12 | Onduline (straight) | Yes | | | 28 |
| 13 | Onduline (perturbed) | Yes | | | 28 |
| 14 | Hierarchical WFSensing | | | | 29 |
| 15 | Global MCAO with Very Linear (VL)-WFS | Yes | | | 31 |
| 16 | Dark WFS: Pyramid with dark pin | | | | 37 |
| 17 | Dark WFS: Double arms interferometer | | | | 37 |
| 18 | Dark WFS: double Smartt | | | | 37 |
| 19 | z-invariant with four blocks of glasses | | | | 39 |
| 20 | Multiple spatial frequencies WFS | | | | 38 |
| 21 | Dark LGS: a wire WFS | | Yes | | 39 |
| 22 | INGOT: 2 sides | | Yes | | 40 |
| 23 | INGOT: 3 sides | | Yes | | 40 |
| 24 | INGOT: 4 sides (1st arrangement) | | Yes | | 40 |
| 25 | INGOT: 4 sides (2nd arrangement) | | Yes | | 40 |
| 26 | INGOT: 5 sides | | Yes | | 40 |
| 27 | INGOT: 6 sides | | Yes | | 40 |
| 28 | Planetary disk WFS | Yes | | | 47 |

**Table 1 – Summarizing the nature of the WFSs described in the text.**

While in Table 1 an overall list of the WFS is given, let me express again as this list is somehow a reductionist sample of the various possibilities, and, as originally stated, it only encompass what I had the venture to conceive so it deliberately lack the (several, and some very smart) WFSs that has been devise in the last quarter of a century. The rush to find out the kind of WFS that will become the new standard and will outperform the pyramid WFS is already started and will hopefully soon find a new paradigm. Just to wait for the next one, of course.

## ACKNOWLEDGEMENTS

Thanks are due to all the colleagues and friends that with their work, criticism, and stimulus, contributed to the conception of these and other ideas in the field of Adaptive Optics. Further to the groups that I had the honor to deal with in Padova, Arcetri and Heidelberg, and the whole French school that maintained in these years the status of "the" (friendly) competitor, I want to mention here Francois Rigaut that initiated in several occasions, the thoughts, criticisms and suggestions that led to the exploitation of several of the new concepts described in this manuscript.